\documentclass[10pt,journal]{IEEEtran}
\usepackage{amsfonts}
\IEEEoverridecommandlockouts

\ifCLASSINFOpdf
\else
\fi
\usepackage{epstopdf}
\usepackage{multirow}
\usepackage{cite}
\usepackage{url}
\usepackage{stfloats}
\usepackage{color}
\usepackage{epsfig}
\usepackage{graphicx}
\usepackage{subfigure}
\usepackage{psfig}
\usepackage{epsf}
\usepackage{slashbox}
\usepackage{float}
\usepackage{amssymb }
\usepackage[cmex10]{amsmath}
\begin{document}
\title{ On the  Outage Performance  of SWIPT Based Three-step Two-way  DF Relay Networks }
\author{\IEEEauthorblockN{Yinghui Ye, Liqin Shi,  Xiaoli Chu,~\IEEEmembership{Senior Member,~IEEE},  Hailin Zhang,~\IEEEmembership{Member,~IEEE}, and Guangyue Lu}\vspace*{0em}\\
\thanks{Copyright (c) 2015 IEEE. Personal use of this material is permitted. However, permission to use this material for any other purposes must be obtained from the IEEE by sending a request to pubs-permissions@ieee.org}
\thanks{Manuscript received XXXX, revised XXXX.  The research reported in this article was supported by the Natural Science Foundation of China
(61801382), the Science and Technology Innovation Team of
Shaanxi Province for Broadband Wireless and Application (2017KCT-30-02).}
\thanks{Yinghui Ye, Liqin Shi, and Hailin Zhang are with the  Xidian University.}
\thanks{Xiaoli Chu is with the Department of Electronic and Electrical Engineering,
The University of Sheffield, U.K. (e-mail: x.chu@sheffield.ac.uk).}
\thanks{Guangyue Lu is with the Xi'an University of Posts \& Telecommunications,
Xi'an, Shaanxi, P. R. China.}
}
\markboth{IEEE Transactions on Vehicular Technology}
{Ye\MakeLowercase{\textit{et al.}}: Outage Performance Analysis of SWIPT Based Three-step Two-way DF relays}
\maketitle
%\IEEEtitleabstractindextext
%\thanks{The research reported in this article was supported by the Natural Science Foundation of China (61701399, 61501371), the Science and Technology Innovation Team of Shaanxi Province for Broadband Wireless and Application (2017KCT-30-02),  the 111 Project of China (B08038), and the Research Program of Education Bureau of Shaanxi Province (17JK0699).}

\begin{abstract}
%Simultaneous wireless information and power transfer is a viable option to prolong  the lifetime of energy-constrained  networks.
In this paper, we study  the  outage performance of simultaneous wireless information and power transfer (SWIPT) based three-step two-way decode-and-forward (DF) relay networks, where both  power-splitting (PS)  and  \lq\lq harvest-then-forward\rq\rq  $ $ are employed. In particular, we derive the  expressions of  terminal-to-terminal (T2T) and system outage probabilities based on a Gaussian-Chebyshev quadrature approximation, and obtain the  T2T and system outage capacities.  The effects of various system parameters, {\color{black} e.g., the static power allocation ratio at the relay,}  symmetric PS, as well as  asymmetric PS, on the outage performance of the investigated network  are examined. It is  shown that our derived expression for T2T outage capacity is more accurate than existing analytical results, and that the asymmetric PS  achieves a higher system outage capacity than the symmetric one when the
 channels between the relay node
and the terminal nodes have different statistic gains.
\end{abstract}
 %¿Õ³öÒ»ÐÐÃüÁî

\IEEEpeerreviewmaketitle
%\vspace{-10pt}
\section{Introduction}
\IEEEPARstart{S}{imultaneous} wireless information and power transfer (SWIPT) has been   recognized as a promising technology to alleviate  the energy constraint  in relay networks  while maintaining reliable communications through a power-splitting (PS)  or time-switching (TS) scheme \cite{7744827,8337780}.
%It allows an energy-constrained node to split or switch the radio frequency (RF) signal in the power or time domain.
 In a bidirectional transmission network, two-way relaying operates in two steps or three steps instead of four steps as required  in one-way relaying,  thus achieving  a higher capacity \cite{T31}. In this regard, {\color{black}SWIPT based three-step (or two-step) two-way relay networks (TWRNs) are receiving increasing attention from the wireless industry and academia. Due to the limited space and our attention on SWIPT based three-step  TWRNs, our work omits the literature review of  SWIPT based two-step TWRNs, which can be referred to the art-of-the-state references \cite{8368152,zhong2018outage,7831382}}.

As the low complexity of hardware is very important to  energy-constrained networks, while the circuit of three-step TWRNs is  simper than that of  two-step TWRNs,
 three-step TWRNs have attracted
vast attention recently \cite{T31,EL,8287997,8361446, 2017CL, jiang2017outage, 8377371,8364583}.
%In \cite{T31}, the expressions for terminal-to-terminal outage probability have been derived to design wireless power transfer schemes in TS based three-step AF TWRNs.
The authors of \cite{T31} derived the terminal-to-terminal (T2T)
outage probability, which considers the outage events at the terminal nodes
independently, for three wireless power transfer schemes
in TS based three-step amplify-and-forward (AF) TWRNs.
Considering  three-step AF multiplicative TWRNs,  the authors in \cite{EL} proposed  a symmetric PS scheme\footnote{In PS based three-step TWRNs, the relay is equipped with one power splitter and  the PS ratios for both terminal-relay links may be unequal. If the PS ratios for two terminal-relay links  are always equal, we refer to the PS scheme as the symmetric PS; otherwise, we refer to it as the asymmetric PS.} and investigated the T2T outage probability. To minimize the system outage probability, an optimal symmetric PS scheme and  an optimal asymmetric scheme were  developed in \cite{8287997} and \cite{8361446}, respectively.
Recently,  three-step  two-way relaying has been extended to  DF  relaying  \cite{2017CL}, where both upper and lower bounds of the T2T outage capacity were obtained for  the symmetric PS scheme.
%However, there are  large gaps between the theoretical bounds and simulation results \cite{2017CL}.
The authors of \cite{jiang2017outage} derived  expressions of system outage probability  for three-step  TWRNs with TS(PS) SWIPT.
 To explore the advantages of both AF and DF relaying protocols, the authors studied the T2T performance of a SWIPT based three-step TWR with hybrid decode-amplify-forward (HDAF) \cite{8377371}. The combination of SWIPT based three-step with other cognitive radio was reported in \cite{8364583}, aiming to analyze the system outage probability for the primary user and the secondary user.

\begin{figure}
  \centering
  \includegraphics[width=0.35\textwidth]{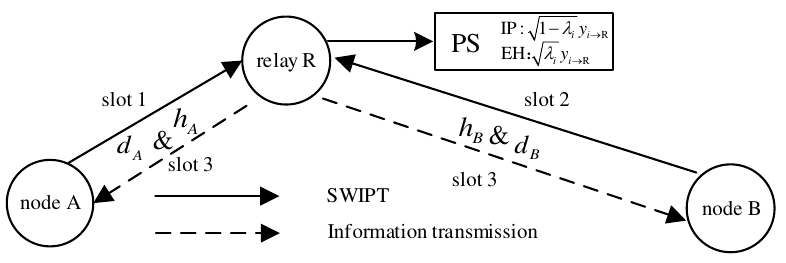}\\
  \caption{System model of the three-step two-way DF relay network.}\label{fig0}
\end{figure}
Although these works mentioned above have laid a solid foundation for the understanding of SWIPT based three-step TWRNs, there are still large gaps needed to be filled. In particular, for a SWIPT based three-step DF TWRN, the existing theoretical expression  of T2T outage capacity \cite{2017CL} is far from the simulation results. Besides, the authors in \cite{jiang2017outage} and \cite{8364583} focused on the analysis of system outage probability,  while they ignored the  channel reciprocity of terminal-relay links  when deriving the expression of system outage probability. Accordingly, the system outage performance of SWIPT based three-step DF TWRNs is still far from being well understood.
Motivated by the above observations, in this paper, we study the T2T and system outage probabilities  of  three-step DF TWRNs, where  PS  and \lq\lq harvest-then-forward\rq\rq $ $ are employed  at the relay node, and derive  expressions for both of them. Different from the T2T outage probability, the system outage probability  evaluates the overall  outage performance of both T2T links (i.e., the two one-way relay links) in the TWRNs,  making the analysis and derivation much  challenging than those of T2T outage performance.
Based on the derived expressions, the impacts of various system parameters, such as the relay location and  the energy conversion  efficiency, on the system outage performance are analyzed. The performance comparison between symmetric and asymmetric  PS schemes is also presented.
\section{System model}
We consider a DF TWRN, where two terminal nodes $\rm{A}$ and $\rm{B}$ exchange information with the aid of an energy-constrained relay node $\rm{R}$ based on  a PS scheme, {\color{black}as shown in Fig. 1.}  The whole block time $T$  has three transmission slots, where $\beta T$ ({\small{$0<\beta<0.5$}}) is the time proportion for the relay to harvest energy  and decode signal from one  terminal node \cite{2017CL}. The \lq\lq harvest-then-forward\rq\rq   $ $ scheme is adopted to encourage the relay to assist two terminal nodes.  All nodes are equipped with a single antenna and operate in a half-duplex mode. We assume that there is no direct link between $\rm{A}$ and $\rm{B}$. {\color{black}Note that our considered network can be applied to many energy-constrained wireless scenarios, e.g., the wireless sensor network  in a toxic environments and the wireless body area network \cite{jiang2017outage}.}

  { The channel} model is given by  {\small{$|h_{i}|^2d_{i}^{ - \alpha }$ ($i = \rm{A} \;\rm{or}\; \rm{B}$}}), where  {\small{${h_i} \sim \mathcal{CN}\left( {0,{\mu_i}} \right)$}} is the $i$$-$$\rm{R}$ channel fading  coefficient; ${d_{ i}}$ is the $i$-$\rm{R}$ distance; $\alpha $ is the path loss exponent. Thus the statistic channel gain of $i$$-$$\rm{R}$ link is given by $\mu_id_{i}^{ - \alpha }$. For analytical tractability, we  further ignore the consumption of  the transceiver circuitry at $\rm{R}$ \cite{2017CL}.

At the first and the second time slots (each of the time slot is $\beta T$),   nodes  $\rm{A}$ and $\rm{B}$ transmit the normalized information signals  $x_{\rm{A}}$  and $x_{\rm{B}}$ to the relay, respectively. The received signal at the relay $\rm{R}$ from terminal node $i$  is
{\small{${y_{ i\to {\rm{R}}}} = {h_{ i}}\sqrt {{P_{ i}}d_{i}^{-\alpha}} {x_{ i}} + {n_{ iR}}$}},
where {\small{${n_{iR}} \sim {\rm{{\cal C}{\cal N}}}\left( {0,\sigma _{iR}^2} \right)$}} is the additive white Gaussian noise (AWGN) and $P_{i}$ denotes the transmit power of terminal node  $i$. Here, we assume that\footnote{\color{black}The assumption will not cause any loss of generality to  our analysis. This is because the average signal-to-noise-ratios (SNRs) of all channels are essential for
the analysis of system outage performance, and  the average SNRs depend on the transmit power at terminals and all the fading channels.  Although $P_A=P_B$, the independent fading channels  will have different values for their parameters, e.g.,  the distance of the terminal-relay link, making the average SNRs of all channels different.} {\small{$P_{\rm{A}}=P_{\rm{B}}=P_0$}}. Using the  PS scheme, the received signal {\small{${y_{ i\to {\rm{R}}}}$}} is split to two parts: {\small{$\sqrt{{\lambda _i}} {y_{ i\to {\rm{R}}}}$}} for energy harvesting {(EH) and {\small{$\sqrt{1-{\lambda _i}} {y_{ i\to {\rm{R}}}}$}} for information processing. Thus, the total received energy  from both  terminal nodes and the received SNR from $i$ to $\rm{R}$ are  given  by
{\small{${E_{{\rm{total}}}} = {P_0}\eta \beta T\left( {{\lambda _{\rm{A}}}|{h_{\rm{A}}}{|^2}d_{\rm{A}}^{ - \alpha } + {\lambda _{\rm{B}}}|{h_{\rm{B}}}{|^2}d_{\rm{B}}^{ - \alpha }} \right)$}} and
{\small{${\gamma _{i \to {\rm{R}}}} = {P _0}{\left| {{h_i}} \right|^2}\left( {1 - {\lambda _i}} \right)d_i^{ - \alpha }\sigma _{i\rm{R}}^{-2}$}}, respectively, where $\eta $ is the energy conversion efficiency and $\lambda_i$ is the PS ratio  of terminal node $i$. Note that $\lambda_{\rm{A}}$ may be different from $\lambda_{\rm{B}}$ in three-step relay networks, which makes it possible to more efficiently use of the received radio frequency (RF) signal at the relay.

Let $\widetilde{x}_{i}$  be the decoded signal for $x_i$. {\color{black} If both $x_{\rm{A}}$ and $x_{\rm{B}}$ are successfully decoded during the first two slots, then at the third  time slot $(1-2\beta) T$ the relay $\rm{R}$ broadcasts the normalized signal ${\small{{x_R} = {\theta _{\rm{A}}}{x_{\rm{A}}} + {\theta _{\rm{B}}}{x_{\rm{B}}}}}$  to both terminal nodes using the harvested energy $E_{\rm{total}}$, where $ {\theta _{\rm{A}}^2 + \theta _{\rm{B}}^2 = 1} $ and $\theta _{i}^2 \left( {i = {\rm{A}},{\rm{B}}} \right)$  is a static\footnote{Note that the system capacity can be further improved if a dynamic power allocation  ratio, which is adjusted based on the instantaneous CSI (or instantaneous channel gain) of terminal-relay links  instead of statistic one, is adopted. This will be considered in our future work.} power allocation ratio determining how the relay allocates the power for the decoded signals based on the statistic channel state information (CSI) (or statistic channel gain) of the $i$$-$$\rm{R}$ link.
Thus, the received signal at terminal node  $i$ is given by
{\small{${y_{{\rm{R}} \to i}} = {h_i}\sqrt {{P_{\rm{R}}}d_i^{ - \alpha }} {x_{\rm{R}}} + {n_{{\rm{R}}i}}\mathop {\rm{ = }}\limits^{\left( a \right)} {h_i}\sqrt {{P_{\rm{R}}}d_i^{ - \alpha }} {\theta _{\bar i}}{\tilde x_{\bar i}} + {n_{{\rm{R}}i}}$}},}
where if {\small{$i=\rm{A}$, $\bar i=\rm{B}$; $i=\rm{B}$, $\bar i=\rm{A}$}}; {\small{  ${n_{{\rm{R}}i}} \sim {\rm{{\cal C}{\cal N}}}\left( {0,\sigma _{{\rm{R}}i}^2} \right)$}} is the AWGN; {\small{${P_{\rm{R}}} = \frac{{{E_{\rm{total}}}}}{{\left( {1 - 2\beta } \right)T}}$}} is the transmit power at $\rm{R}$;  {\color{black}step (a) holds under the assumption of   perfect successive interference cancellation (SIC) \cite{2017CL,jiang2017outage,8377371}. This assumption is used here  in order to obtain the upper bound of outage performance for our considered network.} %It is worth mentioning that relaxing this ideal assumption can make our considered network close to the practical scenario, but this is beyond the scope of this paper.}

Accordingly, {\color{black}the SNR of the ${\rm{R}} \to i$ link is given by
{\small{${\gamma _{{\rm{R}} \to i}} = \frac{{{\theta _{\bar i}^2}{\rho _0}\eta \beta }}{{\left( {1 - 2\beta } \right)d_i^\alpha }}|{h_i}{|^2}\left( {{\lambda _{\rm{A}}}|{h_{\rm{A}}}{|^2}d_{\rm{A}}^{ - \alpha } + {\lambda _{\rm{B}}}|{h_{\rm{B}}}{|^2}d_{\rm{B}}^{ - \alpha }} \right)$}},}
where  {\small{${\rho _0} = \frac{{{P_0}}}{{{\sigma ^2}}}$}} and {\small{ $\sigma _{\rm{AR}}^2 = \sigma _{\rm{BR}}^2 = \sigma _{\rm{RA}}^2 = \sigma _{\rm{RB}}^2 ={\sigma ^2}$}}. %Without loss of generality, we assume that $\beta=1/3$.
\section{Outage Probability Analysis}
\subsection{ Terminal-to-Terminal Outage Probability}
%Here, we  derive an analytical expression for the T2T outage probability, which considers the  outage events at two terminals independently and can be used to  evaluate the outage performance of $\rm{A}\to \rm{R} \to \rm{B}$ link and $\rm{B}\to \rm{R} \to \rm{A}$ link.
Let $\mathbb{P} \left(  \cdot  \right)$ denote the probability and $P_{\rm{out}}^{i}$ be the outage probability of  $\bar i\to \rm{R} \to i$ link. % where if {\small{$i=\rm{A}$, $\bar i=\rm{B}$; $i=\rm{B}$, $\bar i=\rm{A}$}}.
For a given target transmission  rate $U$ and the corresponding SNR threshold ${\gamma _{\rm{th}}}=2^U-1$ \cite{2017CL}, $P_{\rm{out}}^{i}$ is written as
%{\small{$P_{{\rm{out}}}^i = 1 -P^{i}_{1}$}},
 \begin{small}
 \begin{align}\label{K1}
 {P_{{\rm{out}}}^i = 1 -P^{i}_{1}}
 \end{align}
 \end{small}where $P^{i}_{1}=\mathbb{P} \left( {{\gamma _{{\rm{R}} \to i}} \ge {\gamma _{{\rm{th}}}},{\gamma _{\bar i \to {\rm{R}}}} \ge {\gamma _{{\rm{th}}}}} \right)$.

%\mathbb{P}\left( {{\gamma _{\overline{i}\to R}} < \gamma _{\rm{th}}} \right)+\mathbb{P}\left( {{\gamma _{R\to i}} < \gamma _{\rm{th}}},{{\gamma _{\overline{i}\to R}} \geq\gamma _{\rm{th}}}\right)\\
%\begin{small}
%\begin{align}\label{K1}
%P_{\rm{out}}^{i}=1-\underbrace {\mathbb{P}\left( {{\gamma _{{\rm{R}}\to i}} \geq \gamma _{\rm{th}}},{{\gamma _{\overline{i}\to {\rm{R}}}} \geq\gamma _{\rm{th}}}\right)}_{P_1^{i}}.
%\end{align}
%\end{small}
Based on  the expression of   $\gamma _{\rm{R}} \to i$ and ${\gamma _{\bar i \to {\rm{R}}}}$, ${P_1^{i}}$ can be rewritten as
\begin{small}
\begin{align}\notag\label{K2}
{P_1^i}&=\mathbb{P}\left[|{h_{\bar i}}{|^2}\geq \max\left(\Phi_{\bar i},\Psi_{\bar i}\right)\right]\\
&=\mathbb{P}\left[|{h_{\bar i}}{|^2}\geq \Psi_{{\bar i}},\Psi_{{\bar i}}\geq\Phi_{{\bar i}}\right]+\mathbb{P}\left[|{h_{\bar i}}{|^2}\geq \Phi_{{\bar i}},\Phi_{{\bar i}}>\Psi_{{\bar i}}\right]
\end{align}
\end{small}where $\Phi_{\bar i}={\frac{{{\gamma _{{\rm{th}}}}d_{\bar i}^\alpha }}{{{\rho _0}\left( {1 - {\lambda _{\bar i}}} \right)}}}$, $\Psi_{{\bar i}}={\frac{{{{{\gamma _{{\rm{th}}}}} \mathord{\left/
 {\vphantom {{{\gamma _{{\rm{th}}}}} {{X_{i}} - {\lambda _{ i}}|{h_{ i}}{|^4}d_{i}^{ - \alpha }}}} \right.
 \kern-\nulldelimiterspace} {{X_{ i}} - {\lambda _{ i}}|{h_{ i}}{|^4}d_{ {i}}^{ - \alpha }}}}}{{{\lambda _{\bar i}}|{h_{ i}}{|^2}}}d_{\bar i}^\alpha }$ and {\color{black}${X_i} = \frac{{{\theta _{\bar i}^2}{\rho _0}\eta \beta }}{{\left( {1 - 2\beta } \right)d_i^\alpha }}$}.

 When $\Psi_{{\bar i}}\geq\Phi_{{\bar i}}$, we have the following inequality
 \begin{small}
 \begin{align}\label{K3}
{a_i}|{h_i}{|^4} + {b_{\bar i}}|{h_i}{|^2} - {c_i} \leq 0
 \end{align}
\end{small}where ${a_i} = {\lambda _i}d_i^{ - \alpha },{b_i} = \frac{{{\gamma _{{\rm{th}}}}{\lambda _i}}}{{{\rho _0}\left( {1 - {\lambda _i}} \right)}}$ and ${c_i} = {\gamma _{{\rm{th}}}}/{X_i}$.

Combining (\ref{K3}) with $|{h_i}{|^2}\geq 0$, the range of $|{h_i}{|^2}$  is $0\leq|{h_i}{|^2}\leq\Omega_i$, where $\Omega_i=\frac{{\sqrt {b_{\bar i}^2 + 4{a_i}{c_i}}  - {b_{\bar i}}}}{{2{a_i}}}$.

 When $\Phi_{{\bar i}}>\Psi_{{\bar i}}$, we have $|{h_i}{|^2}>\Omega_i$.
Thus, ${P_1^i}$ can be calculated as
\begin{small}
\begin{align}\notag\label{K4}
&{P_1^{i}}\!=\!\mathbb{P}\left[|{h_{\bar i}}{|^2}\geq \Phi_{{\bar i}},|{h_i}{|^2}>\Omega_i\right]
\!+\!\mathbb{P}\left[|{h_{\bar i}}{|^2}\geq \Psi_{{\bar i}},0\leq|{h_i}{|^2}\leq\Omega_i \right]\\
&=\exp \left( { - \frac{{{\Phi _{\bar i}}}}{{{\mu _{\bar i}}}} - \frac{{{\Omega _i}}}{{{\mu _i}}}} \right)+\frac{1}{{{\mu _i}}}\int_0^{{\Omega_i}} {\exp \left( { - \frac{{{\Psi_{\bar i}}\left( y \right)}}{{{\mu_{\bar i}}}} - \frac{y}{{{\mu _i}}}} \right)} dy
\end{align}
\end{small}where the second equality holds from  ${\left| {{h_i}} \right|^2} \sim \exp \left( {\frac{1}{{\mu _i}}} \right)$ for $i \in \left\{ {\rm{A},\rm{B}} \right\}$ and $y=|h_i|^2$.

{\color{black}Since it is difficult to find the  closed-form expression for ${P_1^i}$ due to the integral $\int_{{s_1}}^{{s_2}} {\exp ({z_1}x + \frac{{{z_2}}}{x})} dx$ for any value of $z_{1}$ and $z_{2}\neq 0$, we employ Gaussian-Chebyshev
quadrature \cite{8337780,7445146} to obtain an approximation for ${P_1^i}$},  as follows
\begin{small}
\begin{align}\notag\label{K5}
{P_1^i}&\approx \exp \left( { - \frac{{{\Phi _{\bar i}}}}{{{\mu _{\bar i}}}} - \frac{{{\Omega_i}}}{{{\mu _i}}}} \right) + \frac{{\pi {\Omega_i}}}{{2N{\mu _i}}}\sum\limits_{n = 1}^N {\sqrt {1 - \nu _n^2} } \\
&\times\exp \left( { - \frac{{{\Psi _{\bar i}}\left( {\chi _n^{(01)}} \right)}}{{{\mu _{\bar i}}}} - \frac{{\chi _n^{(01)}}}{{{\mu _i}}}} \right)
\end{align}
\end{small}where $N$ is a parameter that determines the tradeoff between complexity and accuracy  for the Gaussian-Chebyshev quadrature based approximation; ${\nu_n} = \cos \frac{{2n - 1}}{{2N}}\pi $, and $\chi _n^{(01)} = \frac{{{\Omega_i }}}{2}{\nu _n} + \frac{{{\Omega_i }}}{2}$. {\color{black}Note that an acceptable accuracy can be achieved for a small value of $N$}, which is verified in our simulations.

Submitting (\ref{K5}) into (\ref{K1}), we obtain the T2T outage probability for  $\bar i\to {\rm{R}} \to i$ link, $P_{{\rm{out}}}^{i}$.

Thus, the outage capacity $\tau _{{\rm{out}}}^{i}$ can be calculated as
\begin{small}
\begin{align}\label{y5}
\tau _{{\rm{out}}}^{i} = \left( {1 - P_{{\rm{out}}}^{i}} \right)U\beta T.
\end{align}
\end{small}

%Thus, the value of $P_{\rm{out}}^{B}$ can be determined by computing $1-{P_1^B}$.
%Similarly, the end-to-end outage probability at node $A$,$P_{\rm{out}}^{A}$, is given by
%\begin{small}
%\begin{align}\notag\label{K6}
%P_{\rm{out}}^{A}&\approx 1-\exp \left( { - \frac{{{\Phi _B}}}{{{\mu _B}}} - \frac{{{\Omega _A}}}{{{\mu _A}}}} \right) - \frac{{\pi {\Omega _A}}}{{2N{\mu _A}}}\sum\limits_{n = 1}^N {\sqrt {1 - \nu _n^2} } \\
%&\times\exp \left( { - \frac{{{\Psi _B}\left( {\chi _n^{(02)}} \right)}}{{{\mu _B}}} - \frac{{\chi _n^{(02)}}}{{{\mu _A}}}} \right)
%\end{align}
%\end{small}where $\Omega_{A}=\frac{{\sqrt {b_B^2 + 4{a_A}{c_A}}  - {b_B}}}{{2{a_A}}}$ and $\chi _n^{(02)} = \frac{{{\Omega _A}}}{2}{\nu _n} + \frac{{{\Omega _A}}}{2}$.
\vspace{-20pt}
\subsection{System Outage Probability}
%{\color{black}Here, the system outage probability of the  SWIPT based three-step two-way DF relay network is studied}.
The system outage probability\footnote{{\color{black}Note that,  compared with traditional three-step TWRNs without SWIPT, the derivation of system outage probability in SWIPT based three-step TWRNs is more challenging since the transmit power of the relay depends on all the channel gains of  terminal-relay links.} } is defined as the probability that either the SNR of $\bar i \to i$ link is less than the SNR  threshold ${\gamma _{\rm{th}}}$.
Thus, the system outage probability is written as
\begin{small}
\begin{align}\label{B1}
P_{{\rm{out}}}^{\rm{S}}=1- \mathbb{P}\left({\mathop  \cup \limits_{i = {\rm{A}},{\rm{B}}} {\gamma _{i \to {\rm{R}}}} \ge {\gamma _{{\rm{th}}}},{\gamma _{{\rm{R}} \to i}} \ge {\gamma _{{\rm{th}}}}}\right)=1-{P^{\rm{S}}_{1}}
\end{align}
\end{small}where {\small{$P^{\rm{S}}_{1}=1-\mathbb{P}({\mathop  \cup \limits_{i = {\rm{A}},{\rm{B}}}|{h_i}{|^2} \ge \max (\Phi_{i}, \Psi_{i})} )$}}.
Based on the expressions of $\Phi_{i}$ and $\Psi_{i}$ defined in \eqref{K2}, $P_{1}^{\rm{S}}$ is rewritten as
 \begin{small}
 \begin{align}\label{B2}
 P^{\rm{S}}_{1}=P^{s}_{11}+P^{s}_{12}+P^{s}_{13}+P^{s}_{14}
 \end{align}
 \end{small}where
 \begin{small}
 \begin{align}\notag
 &P^{s}_{11}=\mathbb{P}\left( {|{h_{\rm{A}}}{|^2} \ge {\Psi _A},|{h_{\rm{B}}}{|^2} \ge {\Phi _{\rm{B}}}},{{{\Psi _{\rm{A}}} \ge {\Phi _{\rm{A}}},{\Psi _{\rm{B}}} < {\Phi _{\rm{B}}}}}  \right)\\ \notag
 &P^{s}_{12}=\mathbb{P}\left( {|{h_{\rm{B}}}{|^2} \ge {\Psi _{\rm{B}}},|{h_{\rm{A}}}{|^2} \ge {\Phi _{\rm{A}}}},{{{\Psi _{\rm{A}}} < {\Phi _{\rm{A}}},{\Psi _{\rm{B}}} \ge {\Phi _{\rm{B}}}}} \right) \\ \notag
 &P^{s}_{13}=\mathbb{P}\left( {|{h_{\rm{B}}}{|^2} \ge {\Phi _{\rm{B}}},|{h_{\rm{A}}}{|^2} \ge {\Phi _{\rm{A}}}},{{{\Psi _{\rm{A}}} < {\Phi _{\rm{A}}},{\Psi _{\rm{B}}} < {\Phi _{\rm{B}}}}} \right) \\ \notag
 &P^{s}_{14}=\mathbb{P}\left( {|{h_{\rm{B}}}{|^2} \ge {\Psi _{\rm{B}}},|{h_{\rm{A}}}{|^2} \ge {\Psi _{\rm{A}}}}, {{{\Psi _{\rm{A}}} \geq {\Phi _{\rm{A}}},{\Psi _{\rm{B}}} \geq {\Phi _{\rm{B}}}}} \right).
 \end{align}
 \end{small}
\subsubsection{ $P^{s}_{11}$}
If  {\small{${\Psi _{\rm{A}}} \ge {\Phi _{\rm{A}}}$ and ${\Psi _{\rm{B}}} < {\Phi _{\rm{B}}}$}}, we have
\begin{small}
\begin{align}\label{B3}
\left\{ \begin{array}{l}
{a_{\rm{B}}}|{h_{\rm{B}}}{|^4} + {b_{\rm{A}}}|{h_{\rm{B}}}{|^2} - {c_{\rm{B}}} \le 0\\
{a_{\rm{A}}}|{h_{\rm{A}}}{|^4} + {b_{\rm{B}}}|{h_{\rm{A}}}{|^2} - {c_{\rm{A}}} > 0
\end{array} \right..
\end{align}
\end{small}Since {\small{$|{h_{\rm{A}}}{|^2}\geq 0$}} and {\small{$|{h_{\rm{B}}}{|^2}\geq 0$}}, the solution to \eqref{B3} is given by {\small{$0 \le |{h_{\rm{B}}}{|^2} \le \Omega_{B}$}} and {\small{$|{h_{\rm{A}}}{|^2} > \Omega_{{\rm{A}}}$}}, where {\small{$\Omega_i=\frac{{\sqrt {b_{\bar i}^2 + 4{a_i}{c_i}}  - {b_{\bar i}}}}{{2{a_i}}}$, $i={\rm{A}},{\rm{B}}$}}.
%Let $\Omega_{A}=\frac{{\sqrt {b_B^2 + 4{a_A}{c_A}}  - {b_B}}}{{2{a_A}}}$.
Then $P^{s}_{11}$ can be computed as
\begin{footnotesize}
\begin{align}\label{B4}
&P^{s}_{11}%=\mathbb{P}\left( |{h_{\rm{A}}}{|^2} \ge \max \left( {{\Psi _{\rm{A}}},{\Omega _{\rm{A}}}} \right),{\Phi _{\rm{B}}} \le |{h_{\rm{B}}}{|^2} \le {\Omega _{\rm{B}}} \right)\\ \notag
=\frac{1}{{{\mu _{\rm{B}}}}}\int_{{\Phi _{\rm{B}}}}^{\max \left( {{\Phi _{\rm{B}}},{\Omega _{\rm{B}}}} \right)} {\exp \left( { - \frac{{\phi_{{\rm{A}}} \left( x \right)}}{{{\mu _{\rm{A}}}}} - \frac{x}{{{\mu _{\rm{B}}}}}} \right)} dx\\ \notag
&\approx\frac{{\pi (\max \left( {{\Phi _{\rm{B}}},{\Omega _{\rm{B}}}} \right) - {\Phi _{\rm{B}}})}}{{2N{\mu _{\rm{B}}}}}\sum\limits_{n = 1}^N  \sqrt {1 - \nu_n^2} \exp \!\!\left( \!\!{ - \frac{{\phi_{\rm{A}} \left( {\chi_n^{(1)}} \right)}}{{{\mu_{\rm{A}}}}} \!- \!\frac{{\chi_n^{(1)}}}{{{\mu _{\rm{B}}}}}} \right)
\end{align}
\end{footnotesize}where the approximation is obtained based on  Gaussian-Chebyshev quadrature, $\phi_{\rm{A}} \left( x \right) = \max \left( {{\Psi _{\rm{A}}},{\Omega _{\rm{A}}}} \right)$,  and $\chi_n^{(1)} = \frac{{(\max \left( {{\Phi _{\rm{B}}},{\Omega _{\rm{B}}}} \right) - {\Phi _{\rm{B}}})}}{2}{\nu _n} + \frac{{(\max \left( {{\Phi _{\rm{B}}},{\Omega _{\rm{B}}}} \right) + {\Phi _{\rm{B}}})}}{2}$.

%Since it is difficult to find the accurate closed-form expression for $P^{s}_{11}$ due to the integral $\int_{{s_1}}^{{s_2}} {\exp ({z_1}x + \frac{{{z_2}}}{x})} dx$ with any value of $z_{1}$ and $z_{2}\neq 0$, we employ Gaussian-Chebyshev
%quadrature to obtain an approximation for $P^{s}_{11}$, given by
%\begin{small}
%\begin{align}\notag\label{B5}
%&P^{s}_{11}\approx\\
%&\frac{{\pi (\max \left( {{\Phi _B},{\Omega _B}} \right) - {\Phi _B})}}{{2N{\mu _B}}}\sum\limits_{n = 1}^N  \sqrt {1 - \nu_n^2} \exp \left( { - \frac{{\phi_{A} \left( {\chi_n^{(1)}} \right)}}{{{\mu_A}}} - \frac{{\chi_n^{(1)}}}{{{\mu _B}}}} \right)
%\end{align}
%\end{small}where $N$ is a parameter that determines the tradeoff between complexity and accuracy, ${\nu_n} = \cos \frac{{2n - 1}}{{2N}}\pi $, and $\chi_n^{(1)} = \frac{{(\max \left( {{\Phi _B},{\Omega _B}} \right) - {\Phi _B})}}{2}{\nu _n} + \frac{{(\max \left( {{\Phi _B},{\Omega _B}} \right) + {\Phi _B})}}{2}$.
\subsubsection{ $P^{s}_{12}$}
If {\small{${\Psi _{\rm{A}}} < {\Phi _{\rm{A}}}$}} and {\small{${\Psi _B} \geq {\Phi _{\rm{B}}}$}}, we have {\small{$0 \le |{h_{\rm{A}}}{|^2} \le {\Omega _{\rm{A}}}$}} and {\small{$|{h_{\rm{B}}}{|^2} > {\Omega _{\rm{B}}}$}}.
Similar to $P^{s}_{11}$, $P^{s}_{12}$ can be calculated as
\begin{footnotesize}
\begin{align}\label{B6}\notag
P^{s}_{12}&\approx%=\mathbb{P}\left( |{h_{\rm{B}}}{|^2} \ge \max \left( {{\Psi _{\rm{B}}},{\Omega _{\rm{B}}}} \right),{\Phi _{\rm{A}}} \le |{h_{\rm{A}}}{|^2} \le {\Omega _{\rm{A}}} \right)\approx\\
 \frac{{\pi (\max \left( {{\Phi _{\rm{A}}},{\Omega _{\rm{A}}}} \right) - {\Phi _{\rm{A}}})}}{{2N{\mu _{\rm{A}}}}}\\
 &\times \sum\limits_{n = 1}^N  \sqrt {1 - \nu _n^2} \exp \left( { - \frac{{{\phi _{\rm{B}}}\left( {\chi _n^{(2)}} \right)}}{{{\mu _{\rm{B}}}}} - \frac{{\chi _n^{(2)}}}{{{\mu _{\rm{A}}}}}} \right)
\end{align}
\end{footnotesize}where {\small{$\phi_{\rm{B}} \left( x \right) = \max \left( {{\Psi _{\rm{B}}},{\Omega _{\rm{B}}}} \right)$}} with {\small{${{{\left| {{h_{\rm{A}}}} \right|}^2}}=x$}} and {\small{$\chi_n^{(2)} = \frac{{(\max \left( {{\Phi _{\rm{A}}},{\Omega _{\rm{A}}}} \right) - {\Phi _{\rm{A}}})}}{2}{\nu _n} + \frac{{(\max \left( {{\Phi _{\rm{A}}},{\Omega _{\rm{A}}}} \right) + {\Phi _{\rm{A}}})}}{2}$}}.
\subsubsection{$P^{s}_{13}$}
If   ${\Psi _{\rm{A}}} < {\Phi _{\rm{A}}}$ and ${\Psi _{\rm{B}}} < {\Phi _{\rm{B}}}$, the ranges of $|{h_{\rm{B}}}{|^2}$ and $|{h_{\rm{A}}}{|^2}$ are determined by $|{h_{\rm{B}}}{|^2}>\Omega_{\rm{B}}$ and $|{h_{\rm{A}}}{|^2}>\Omega_{\rm{A}}$, respectively. Combining these ranges with $|{h_{\rm{B}}}{|^2} \ge {\Phi _{\rm{B}}}$ and $|{h_{\rm{A}}}{|^2} \ge {\Phi _{\rm{A}}}$, $P^{s}_{13}$ can be computed as
\begin{small}
\begin{align}\notag\label{B7}
P^{s}_{13}&=\mathbb{P}\left( |{h_{\rm{B}}}{|^2} \ge \max \left( {{\Phi _{\rm{B}}},{\Omega _{\rm{B}}}} \right),|{h_{\rm{A}}}{|^2} \ge \max \left( {{\Phi _{\rm{A}}},{\Omega _{\rm{A}}}} \right) \right)\\
&=\exp \left( { - \frac{{\max \left( {{\Phi _{\rm{A}}},{\Omega _{\rm{A}}}} \right)}}{{{\mu _{\rm{A}}}}} - \frac{{\max \left( {{\Phi _{\rm{B}}},{\Omega _{\rm{B}}}} \right)}}{{{\mu _{\rm{B}}}}}} \right).
\end{align}
\end{small}
\subsubsection{ $P^{s}_{14}$}
If ${\Psi _{\rm{A}}} \geq {\Phi _{\rm{A}}}$ and ${\Psi _{\rm{B}}} \geq {\Phi _{\rm{B}}}$, both $0\leq|{h_{\rm{B}}}{|^2}\leq\Omega_{\rm{B}}$ and $0\leq|{h_{\rm{A}}}{|^2}\leq\Omega_{\rm{A}}$ should be satisfied. Then $P^{s}_{14}$ can be denoted as
\begin{small}
\begin{align}\label{B8}
P^{s}_{14}=\mathbb{P}\left({\Psi _{\rm{A}}} \le |{h_{\rm{A}}}{|^2} \le {\Omega _{\rm{A}}},{\Psi _{\rm{B}}} \le |{h_{\rm{B}}}{|^2} \le {\Omega _{\rm{B}}}\right).
\end{align}
\end{small}Note that $P^{s}_{14}=0$ always holds for the case with ${\Psi _{\rm{A}}}> {\Omega _{\rm{A}}}$ or ${\Psi _{\rm{B}}}> {\Omega _{\rm{B}}}$.
Since ${\Psi _{\rm{A}}}$ and ${\Psi _{\rm{B}}}$ are highly correlated, it is difficult to derive $P^{s}_{14}$.
%Nevertheless,  we have found the integral of region of $P^{s}_{14}$ and determined the value of $P^{s}_{14}$ by computing the integral value over the region in what follows.
Let $|{h_{\rm{A}}}{|^2} = x$ and $|{h_{\rm{B}}}{|^2} = y$. We find that  the integral  region for calculating $P^{s}_{14}$ is bounded by the following  4 lines: $y = \frac{{{\gamma _{{\rm{th}}}}/{X_{\rm{A}}} - {\lambda _{\rm{A}}}{x^2}d_{\rm{A}}^{ - \alpha }}}{{{\lambda _{\rm{B}}}x}}d_{\rm{B}}^\alpha $, $x = \frac{{{\gamma _{{\rm{th}}}}/{X_{\rm{B}}} - {\lambda _{\rm{B}}}{y^2}d_{\rm{B}}^{ - \alpha }}}{{{\lambda _{\rm{A}}}y}}d_{\rm{A}}^\alpha $, $x = {\Omega _{\rm{A}}}$ and $y = {\Omega _{\rm{B}}}$.

Let $(x_1,y_{\Delta})$ denote the intersection of the lines $x = {\Omega _{\rm{A}}}$ and $x = \frac{{{\gamma _{{\rm{th}}}}/{X_{\rm{B}}} - {\lambda _{\rm{B}}}{y^2}d_{\rm{B}}^{ - \alpha }}}{{{\lambda _{\rm{A}}}y}}d_{\rm{A}}^\alpha $. Then $(x_1,y_{\Delta})$ is given by
\begin{small}
\begin{align}\label{B9}
{{x_1} = {\Omega _{\rm{A}}},{y_\Delta } = \frac{{\sqrt {\lambda _{\rm{A}}^2x_1^2 + 4{a_{\rm{B}}}{c_{\rm{B}}}d_{\rm{A}}^{2\alpha }}  - {\lambda _{\rm{A}}}{x_1}}}{{2{a_{\rm{B}}}d_{\rm{A}}^\alpha }}}.
\end{align}
\end{small}Similarly, the intersection of the lines $y = {\Omega _{\rm{B}}}$ and $y = \frac{{{\gamma _{{\rm{th}}}}/{X_{\rm{A}}} - {\lambda _{\rm{A}}}{x^2}d_{\rm{A}}^{ - \alpha }}}{{{\lambda _{\rm{B}}}x}}d_{\rm{B}}^\alpha $, $(x_{\Delta},y_1)$, is given by
\begin{small}
\begin{align}\label{B10}
{{x_\Delta } = \frac{{\sqrt {\lambda _{\rm{B}}^2y_1^2 + 4{a_{\rm{A}}}{c_{\rm{A}}}d_{\rm{B}}^{2\alpha }}  - {\lambda _{\rm{B}}}{y_1}}}{{2{a_{\rm{A}}}d_{\rm{B}}^\alpha }},{y_1} = {\Omega _{\rm{B}}}}.
\end{align}
\end{small}The intersection of the lines $y = {\Omega _{\rm{B}}}$ and $x = \frac{{{\gamma _{{\rm{th}}}}/{X_{\rm{B}}} - {\lambda _{\rm{B}}}{y^2}d_{\rm{B}}^{ - \alpha }}}{{{\lambda _{\rm{A}}}y}}d_{\rm{A}}^\alpha $ is denoted by  $(q_1, y_1)$, where $q_1=\frac{{{\gamma _{{\rm{th}}}}/{X_{\rm{B}}} - {\lambda _{\rm{B}}}y_1^2d_{\rm{B}}^{ - \alpha }}}{{{\lambda _{\rm{A}}}{y_1}}}d_{\rm{A}}^\alpha $.
The intersection of the lines $x = {\Omega _{\rm{A}}}$ and $y = \frac{{{\gamma _{{\rm{th}}}}/{X_{\rm{A}}} - {\lambda _{\rm{A}}}{x^2}d_{\rm{A}}^{ - \alpha }}}{{{\lambda _{\rm{B}}}x}}d_{\rm{B}}^\alpha $ is denoted by $(x_1, q_2)$, where $q_2=\frac{{{\gamma _{{\rm{th}}}}/{X_{\rm{A}}} - {\lambda _{\rm{A}}}x_1^2d_{\rm{A}}^{ - \alpha }}}{{{\lambda _{\rm{B}}}{x_1}}}d_{\rm{B}}^\alpha $.

\begin{figure}
\begin{tabular}{cc}
\begin{minipage}[t]{0.48\linewidth}
    \includegraphics[width = 1\linewidth]{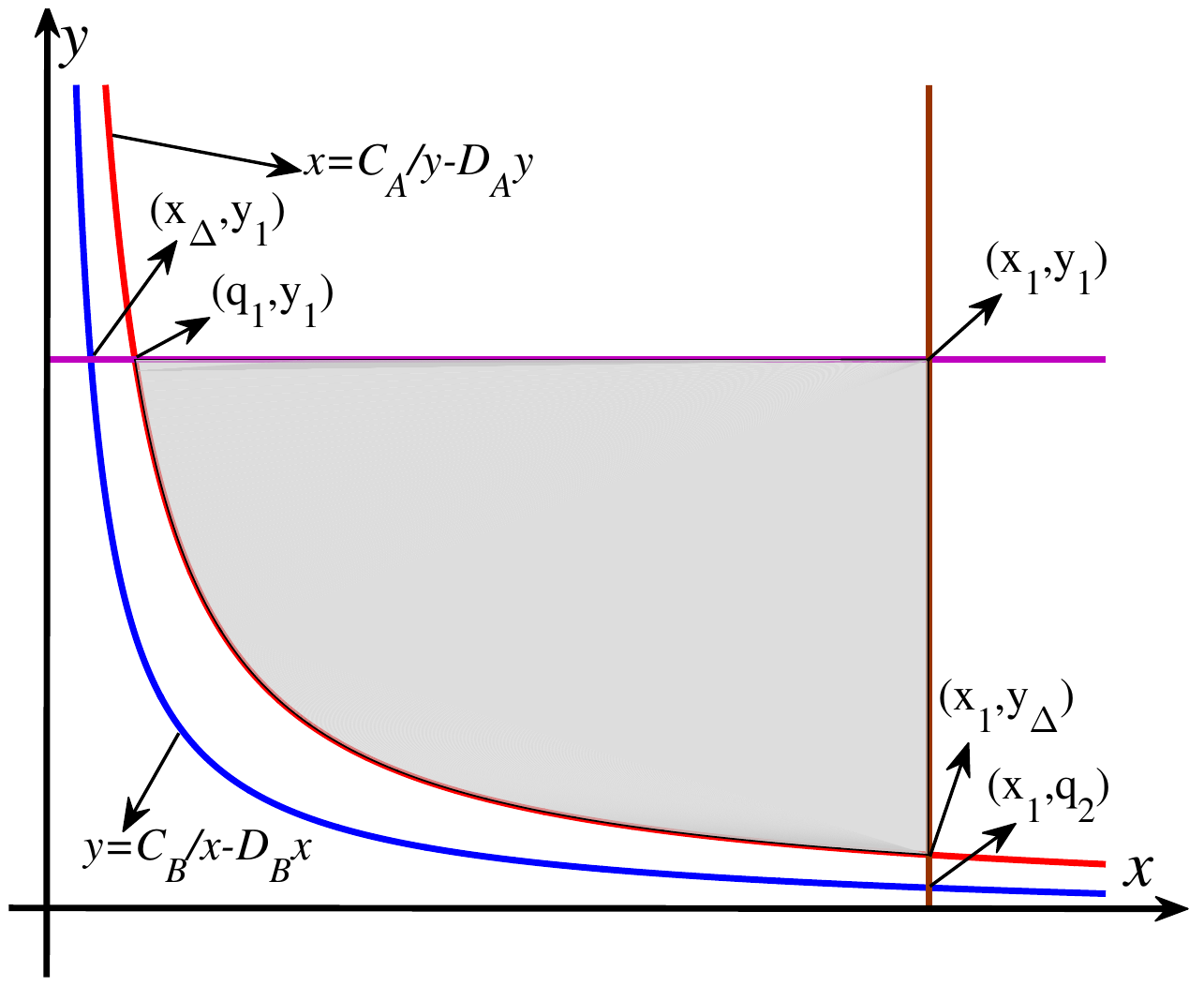}
    \caption{The integral  region for $P^{s}_{14}$ in \textbf{Case II} with $y_{\Delta}\geq q_{2}$.}
\end{minipage}
\begin{minipage}[t]{0.48\linewidth}
    \includegraphics[width = 1\linewidth]{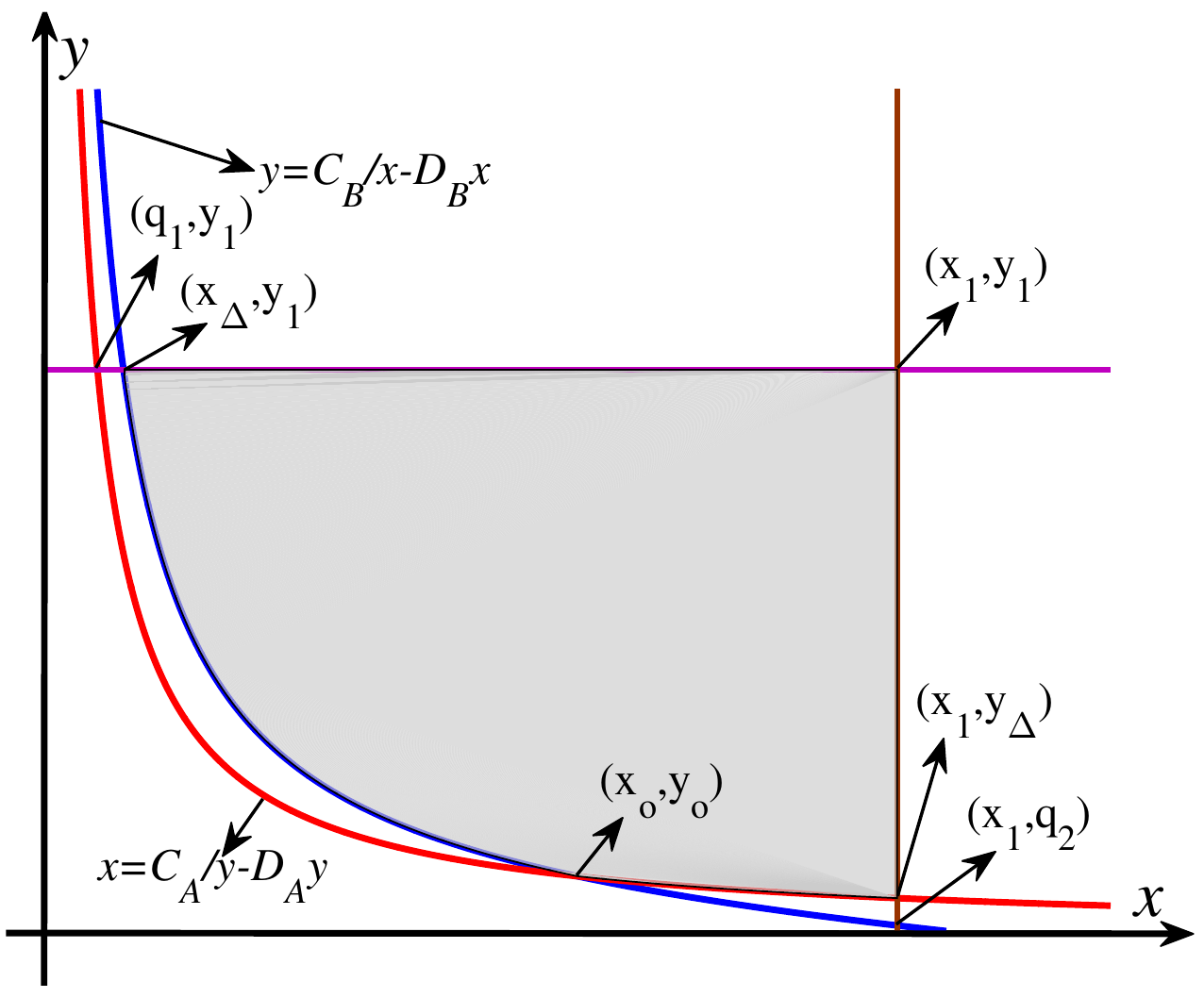}
    \caption{The integral  region for $P^{s}_{14}$ in \textbf{Case III} with $y_{\Delta}\geq q_{2}$.}
\end{minipage}
\end{tabular}
\end{figure}

The intersection of the lines $x = \frac{{{\gamma _{{\rm{th}}}}/{X_{\rm{B}}} - {\lambda _{\rm{B}}}{y^2}d_{\rm{B}}^{ - \alpha }}}{{{\lambda _{\rm{A}}}y}}d_{\rm{A}}^\alpha $ and $y = \frac{{{\gamma _{{\rm{th}}}}/{X_{\rm{A}}} - {\lambda _{\rm{A}}}{x^2}d_{\rm{A}}^{ - \alpha }}}{{{\lambda _{\rm{B}}}x}}d_{\rm{B}}^\alpha $ is determined by
\begin{small}
\begin{align}\label{B12}
{{x = {{{C_{\rm{A}}}} \mathord{\left/
 {\vphantom {{{C_{\rm{A}}}} y}} \right.
 \kern-\nulldelimiterspace} y} - {D_{\rm{A}}}y} \;{\rm{and}}\;y = {{{C_{\rm{B}}}} \mathord{\left/
 {\vphantom {{{C_{\rm{B}}}} x}} \right.
 \kern-\nulldelimiterspace} x} - {D_{\rm{B}}}x}
\end{align}
\end{small}where {\small{${C_i} = {{{\gamma _{{\rm{th}}}}d_i^\alpha } \mathord{\left/
 {\vphantom {{{\gamma _{{\rm{th}}}}d_i^\alpha } {{X_{\bar i}}{\lambda _i}}}} \right.
 \kern-\nulldelimiterspace} {{X_{\bar i}}{\lambda _i}}}$}} and {\small{${D_i} = {{{\lambda _{\bar i}}d_i^\alpha } \mathord{\left/
 {\vphantom {{{\lambda _{\bar i}}d_i^\alpha } {{\lambda _i}d_{\bar i}^\alpha }}} \right.
 \kern-\nulldelimiterspace} {{\lambda _i}d_{\bar i}^\alpha }}$}}. Substituting the {\small{$y = {{{C_{\rm{B}}}} \mathord{\left/
 {\vphantom {{{C_{\rm{B}}}} x}} \right.
 \kern-\nulldelimiterspace} x} - {D_{\rm{B}}}x$}} into {\small{${x = {{{C_{\rm{A}}}} \mathord{\left/
 {\vphantom {{{C_{\rm{A}}}} y}} \right.
 \kern-\nulldelimiterspace} y} - {D_{\rm{A}}}y}$}}, we have
\begin{small}
\begin{align}\label{B13}
 - \left( {{C_{\rm{A}}} + {C_{\rm{B}}}} \right)x^2 + C_{\rm{B}}^2{D_{\rm{A}}} = 0.
\end{align}
\end{small}Since $x>0$, the solution to \eqref{B13} is given by {\small{${x_o} = \sqrt {{{C_{\rm{B}}^2{D_{\rm{A}}}} \mathord{\left/
 {\vphantom {{C_{\rm{B}}^2{D_{\rm{A}}}} {\left( {{C_{\rm{A}}} + {C_{\rm{B}}}} \right)}}} \right.
 \kern-\nulldelimiterspace} {\left( {{C_{\rm{A}}} + {C_{\rm{B}}}} \right)}}} $}}, and the corresponding value of $y$ is
{\small{${y_o} = {{{C_{\rm{B}}}} \mathord{\left/
 {\vphantom {{{C_{\rm{B}}}} {{x_o}}}} \right.
 \kern-\nulldelimiterspace} {{x_o}}} - {D_{\rm{B}}}{x_o}$}}.

According to the values of intersections, the integral  region for $P^{s}_{14}$ can be divided into three cases.

%\begin{figure}
%\subfigure[The integral  region for $P^{s}_{14}$ in \textbf{Case II} with $y_{\Delta}\geq q_{2}$.]{\label fig:a}
%    \includegraphics[width = 0.5\columnwidth]{fig1.pdf}
%    \subfigure[The integral  region for $P^{s}_{14}$ in \textbf{Case III} with $y_{\Delta}\geq q_{2}$.] {\label fig:b}
%    \includegraphics[width = 0.5\columnwidth]{fig2.pdf}
%\end{figure}

\textbf{Case I:} If $\max \left( {{q_1},{x_\Delta }} \right) \ge {x_1}$ or $\max \left( {{q_2},{y_\Delta }} \right) \ge {y_1}$, the integral  region is $0$ and $P^{s}_{14}=0$.

\textbf{Case II:} If $\max \left( {{q_1},{x_\Delta }} \right) < {x_1}$, $\max \left( {{q_2},{y_\Delta }} \right) < {y_1}$, and $x_{o}\leq\max \left( {{q_1},{x_\Delta }} \right)$ (or $x_{o}\geq x_{1}$), the integral  region for $P^{s}_{14}$ is bounded by the following  three lines: $x = {\Omega _{\rm{A}}}$, $y = {\Omega _{\rm{B}}}$, and $y = {{{C_{\rm{B}}}} \mathord{\left/
 {\vphantom {{{C_{\rm{B}}}} x}} \right.
 \kern-\nulldelimiterspace} x} - {D_{\rm{B}}}x $ (or $x = {{{C_{\rm{A}}}} \mathord{\left/
 {\vphantom {{{C_{\rm{A}}}} y}} \right.
 \kern-\nulldelimiterspace} y} - {D_{\rm{A}}}y$).

When $y_{\Delta}\geq q_{2}$, the integral  region for $P^{s}_{14}$ is shown in Fig. 2.
Accordingly, $P^{s}_{14}$ is given by
\begin{small}
\begin{align}\label{B14}
P^{s}_{14}&=\mathbb{P}\left({\Psi _{\rm{A}}} \le |{h_{\rm{A}}}{|^2} \le x_1, y_{\Delta} \le |{h_{\rm{B}}}{|^2} \le y_1\right)\\ \notag
&=\frac{1}{{{\mu _{\rm{B}}}}}\!\!\int_{{y_\Delta }}^{{y_1}} \!\!\!\!{\exp \left( { - \frac{{{C_{\rm{A}}}}}{{{\mu _{\rm{A}}}y}} \!+\! \left( {\frac{{{D_{\rm{A}}}}}{{{\mu _{\rm{A}}}}} - \frac{1}{{{\mu _{\rm{B}}}}}} \right)y} \right)} dy \!-\! {\varepsilon _{\rm{A}}(x_{1},y_{\Delta},y_1)}\\\notag
&\approx \frac{{\pi ({y_1} - {y_\Delta })}}{{2N{\mu _{\rm{B}}}}}\!\!\sum\limits_{n = 1}^N \!\! {\sqrt {1 - \nu _n^2} } \!\exp \left( {\kappa_{\rm{A}} \left( {\chi _n^{(3)}} \right)} \right)\! -\! {\varepsilon _{\rm{A}}(x_{1},y_{\Delta},y_1)}
\end{align}
\end{small}where {\small{${\kappa _i}\left( {\chi _n^{(3)}} \right) =  - \frac{{{C_i}}}{{{\mu _i}\chi _n^{(3)}}} + \left( {\frac{{{D_i}}}{{{\mu _i}}} - \frac{1}{{{\mu _{\bar i}}}}} \right)\chi _n^{(3)}$, $\chi _n^{(3)} = \frac{{({y_1} - {y_\Delta })}}{2}{\nu _n} + \frac{{({y_1} + {y_\Delta })}}{2}$, and ${\varepsilon _i(x_{1},y_{\Delta},y_1)} = \exp \left( { - \frac{{{x_1}}}{{{\mu _i}}}} \right)\left( {\exp \left( { - \frac{{{y_\Delta }}}{{{\mu _{\bar i}}}}} \right) - \exp \left( { - \frac{{{y_1}}}{{{\mu _{\bar i}}}}} \right)} \right)$.}}

Similarly, for the case with $y_{\Delta}< q_{2}$, $P^{s}_{14}$ is given by
\begin{small}
\begin{align}\label{B15}\notag
P^{s}_{14}&\approx \frac{{\pi ({x_1} - {x_\Delta })}}{{2N{\mu _{\rm{A}}}}} \sum\limits_{n = 1}^N\!\! {\sqrt {1 - \nu _n^2} }\! \exp \left( {\kappa_{\rm{B}} \left( {\chi _n^{(4)}} \right)} \right) \\
&- {\varepsilon _{\rm{B}}(y_{1},x_{\Delta},x_1)}
\end{align}
\end{small}where $\chi _n^{(4)} = \frac{{({x_1} - {x_\Delta })}}{2}{\nu _n} + \frac{{({x_1} + {x_\Delta })}}{2}$.

\textbf{Case III:} If $\max \left( {{q_1},{x_\Delta }} \right) < {x_1}$, $\max \left( {{q_2},{y_\Delta }} \right) < {y_1}$, and $\max \left( {{q_1},{x_\Delta }} \right)<x_{o}<x_{1}$, the integral  region for $P^{s}_{14}$ is bounded by four lines.

For the case with $y_{\Delta}\geq q_{2}$, the integral  region for $P^{s}_{14}$ is shown in Fig. 3 and $P^{s}_{14}$ is given by
\begin{small}
\begin{align}\notag\label{B16}
P^{s}_{14}&=\frac{1}{{{\mu _{\rm{A}}}}}\!\int_{{x_\Delta }}^{{x_o}} {\!\!\exp \left( {{\kappa _{\rm{B}}}\left( x \right)} \right)} dx + \frac{1}{{{\mu _{\rm{B}}}}}\!\int_{{y_\Delta }}^{{y_o}} {\!\!\exp \left( {{\kappa _{\rm{A}}}\left( y \right)} \right)} dy \!-\!\Lambda \\ \notag
&\approx \frac{{\pi ({x_o} - {x_\Delta })}}{{2N{\mu _{\rm{A}}}}}\sum\limits_{n = 1}^N {\sqrt {1 - \nu _n^2} } \exp \left( {{\kappa _{\rm{B}}}\left( {\chi _n^{(5)}} \right)} \right) \\
&+ \frac{{\pi ({y_o} - {y_\Delta })}}{{2N{\mu _{\rm{B}}}}}\sum\limits_{n = 1}^N {\sqrt {1 - \nu _n^2} } \exp\left( {{\kappa _{\rm{A}}}\left( {\chi _n^{(6)}} \right)} \right) \!-\!\Lambda
\end{align}
\end{small}where {\small$\Lambda={\varepsilon _{\rm{B}}}({y_1},{x_\Delta },{x_o}) \!\!\!- \!\!\!{\varepsilon _{\rm{A}}}({x_1},{y_\Delta },{y_o}) + \varphi ({x_o},{x_1},{y_o},{y_1})$, $\varphi ({x_o},{x_1},{y_o},{y_1})=\left( {\exp \left( { - \frac{{{x_o}}}{{{\mu _{\rm{A}}}}}} \right) - \exp \left( { - \frac{{{x_1}}}{{{\mu _{\rm{A}}}}}} \right)} \right)\left( {\exp \left( { - \frac{{{y_o}}}{{{\mu _{\rm{B}}}}}} \right) - \exp \left( { - \frac{{{y_1}}}{{{\mu _{\rm{B}}}}}} \right)} \right)$, $\chi _n^{(5)} = \frac{{({x_o} - {x_\Delta })}}{2}{\nu _n} + \frac{{({x_o} + {x_\Delta })}}{2}$ and $\chi _n^{(6)} = \frac{{({y_o} - {y_\Delta })}}{2}{\nu _n} + \frac{{({y_o} + {y_\Delta })}}{2}$.}

%\begin{figure}
%  \centering
%  \includegraphics[width=0.35\textwidth]{fig2.pdf}\\
%  \caption{The integral of region for $P^{s}_{14}$ in \textbf{Case III} with $y_{\Delta}\geq q_{2}$.}\label{fig2}
%\end{figure}
For the case with $y_{\Delta}< q_{2}$, $P^{s}_{14}$ is given by
\begin{small}
\begin{align}\label{B17}\notag
&P^{s}_{14}=\frac{1}{{{\mu _{\rm{A}}}}}\int_{{x_o }}^{{x_1}} {\exp \left( {{\kappa _{\rm{B}}}\left( x \right)} \right)} dx + \frac{1}{{{\mu _{\rm{B}}}}}\int_{{y_o }}^{{y_1}} {\exp \left( {{\kappa _{\rm{A}}}\left( y \right)} \right)} dy  \\
&-{\varepsilon _{\rm{B}}}({y_o},{x_o },{x_1}) - {\varepsilon _{\rm{A}}}({x_1},{y_o },{y_1})\\ \notag
&\approx \frac{{\pi ({x_1} - {x_o })}}{{2N{\mu _{\rm{A}}}}}\!\!\sum\limits_{n = 1}^N\!\! {\sqrt {1 - \nu _n^2} }\! \exp \left( {{\kappa _{\rm{B}}}\left( {\chi _n^{(7)}} \right)} \right)\!-\!{\varepsilon _{\rm{B}}}({y_o},{x_o },{x_1})\\ \notag
& +\frac{{\pi ({y_1} - {y_o })}}{{2N{\mu _{\rm{B}}}}}\!\sum\limits_{n = 1}^N \!\!{\sqrt {1 - \nu _n^2} } \!\exp \left( {{\kappa _{\rm{A}}}\left( {\chi _n^{(8)}} \right)} \right)\!- \! {\varepsilon _{\rm{A}}}({x_1},{y_o },{y_1})
\end{align}
\end{small}where {\small{$\chi _n^{(7)} = \frac{{({x_1} - {x_o })}}{2}{\nu _n} + \frac{{({x_o} + {x_1 })}}{2}$ and $\chi _n^{(8)} = \frac{{({y_1} - {y_o })}}{2}{\nu _n} + \frac{{({y_o} + {y_1 })}}{2}$.}}
Substituting \eqref{B4}, \eqref{B6}, \eqref{B7} and \eqref{B14} (or \eqref{B15}, or \eqref{B16}, or \eqref{B17}, or 0) into \eqref{B2} and \eqref{B1},  $P_{\rm{out}}^{\rm{S}}$ is determined.

\begin{figure}
  \centering
  \includegraphics[width=0.32\textwidth]{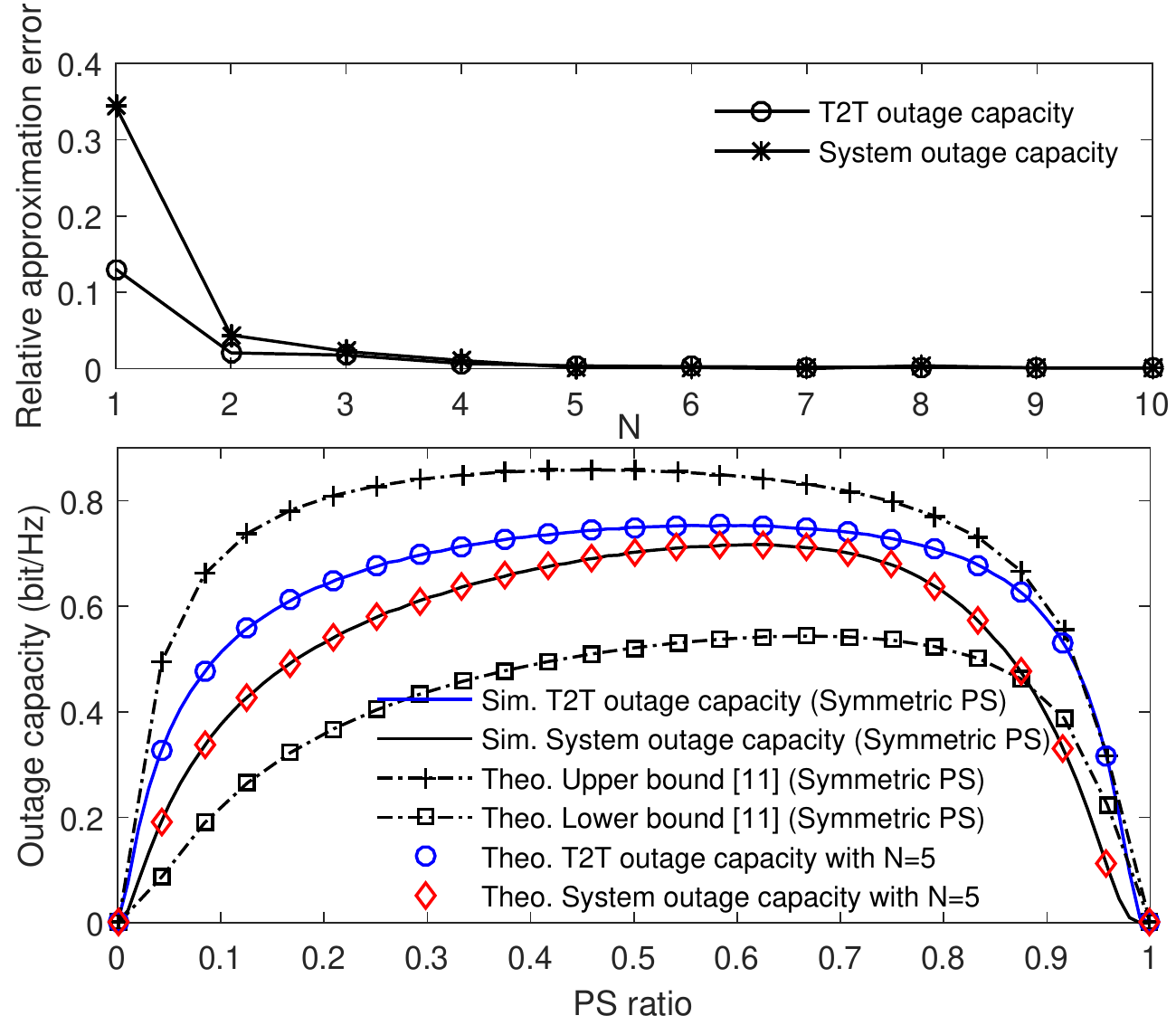}\\
  \caption{The accuracy of our derived results.}
\end{figure}
The system outage capacity $\tau_{\rm{out}}^{\rm{S}}$ is given as
\begin{small}
\begin{align}\label{y6}
\tau_{\rm{out}}^{\rm{S}}=\left(1-P_{\rm{out}}^{\rm{S}} \right)U\beta T.
\end{align}
\end{small}

\emph{Remark 1.} The derived results can serve the following purposes. The first purpose is to obtain an accurate   outage capacity  with a small $N$ instead of the computer simulations. The second purpose is to observe some insights regarding the design of the PS ratios and the static power allocation ratio from the curves obtained by the derived results. In our considered network, the  PS ratios and the static power allocation ratio are designed based on the statistic channel gains instead of instantaneous channel gains, it is practical to obtain the insights  offline regarding the design of PS ratios and the static power allocation ratio and such approach has also adopted in many works, e.g., \cite{7831382}.  A concrete example is Fig. 5 of the revised manuscript, where the optimal system outage
capacity and PS ratios for two PS schemes are obtained by maximizing the derived system outage capacity expression as in problem (23).
It can be observed that  the optimal $\lambda_i$ decreases  with the increase of the statistic channel gain of $i$$-$$\rm{R}$ link.
\begin{align}
\begin{array}{l}
\mathop {\max }\limits_{{\lambda _{\rm{A}}},{\lambda _{\rm{B}}} } \;\tau_{{\rm{out}}}^s\\
s.t.\;\;0 \le {\lambda _{\rm{A}}},{\lambda _{\rm{B}}} \le 1
\end{array}
\end{align}
Lastly, we can obtain the  diversity gain of the considered network, shown in the next subsection.
\begin{figure*}
\begin{minipage}[t]{0.332\linewidth}
\centering
\includegraphics[width=2.33in,height=1.59in]{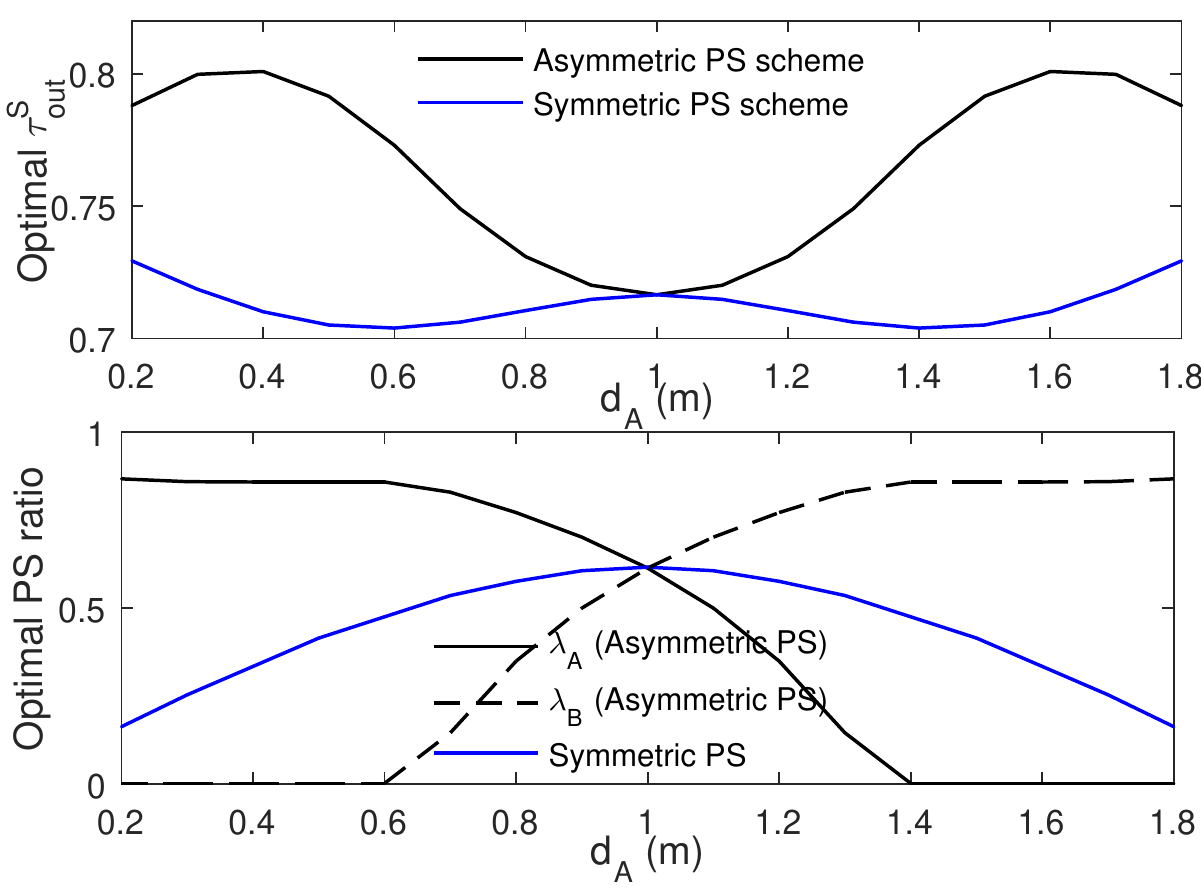}
\centering \caption{Optimal $\tau_{\rm{out}}^{\rm{S}}$ and PS ratio  versus $d_{\rm{A}}$.}
\end{minipage}
\begin{minipage}[t]{0.332\linewidth}
\centering
\includegraphics[width=2.33in,height=1.59in]{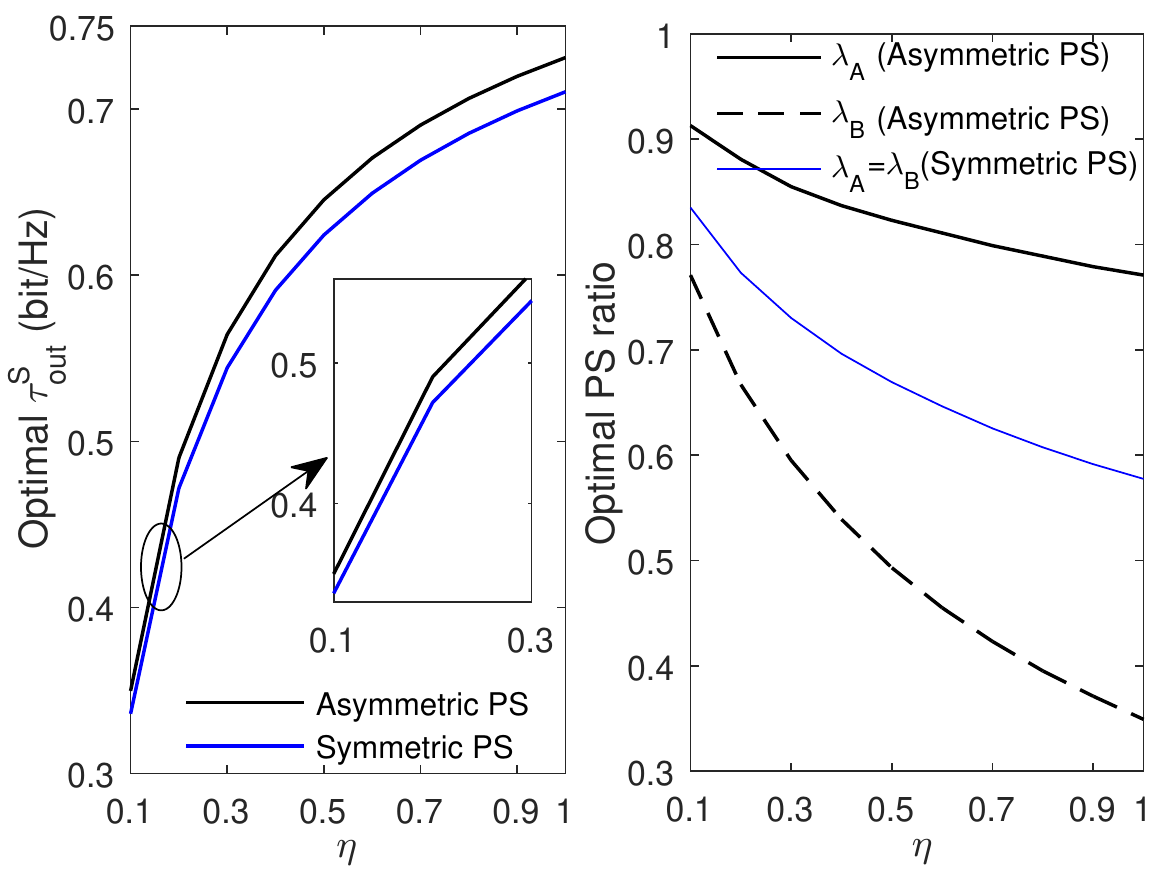}
\centering \caption{ Optimal $\tau_{\rm{out}}^{\rm{S}}$ and PS ratio  versus $\eta$.}
\end{minipage}
\begin{minipage}[t]{0.332\linewidth}
\centering
\includegraphics[width=2.33in,height=1.59in]{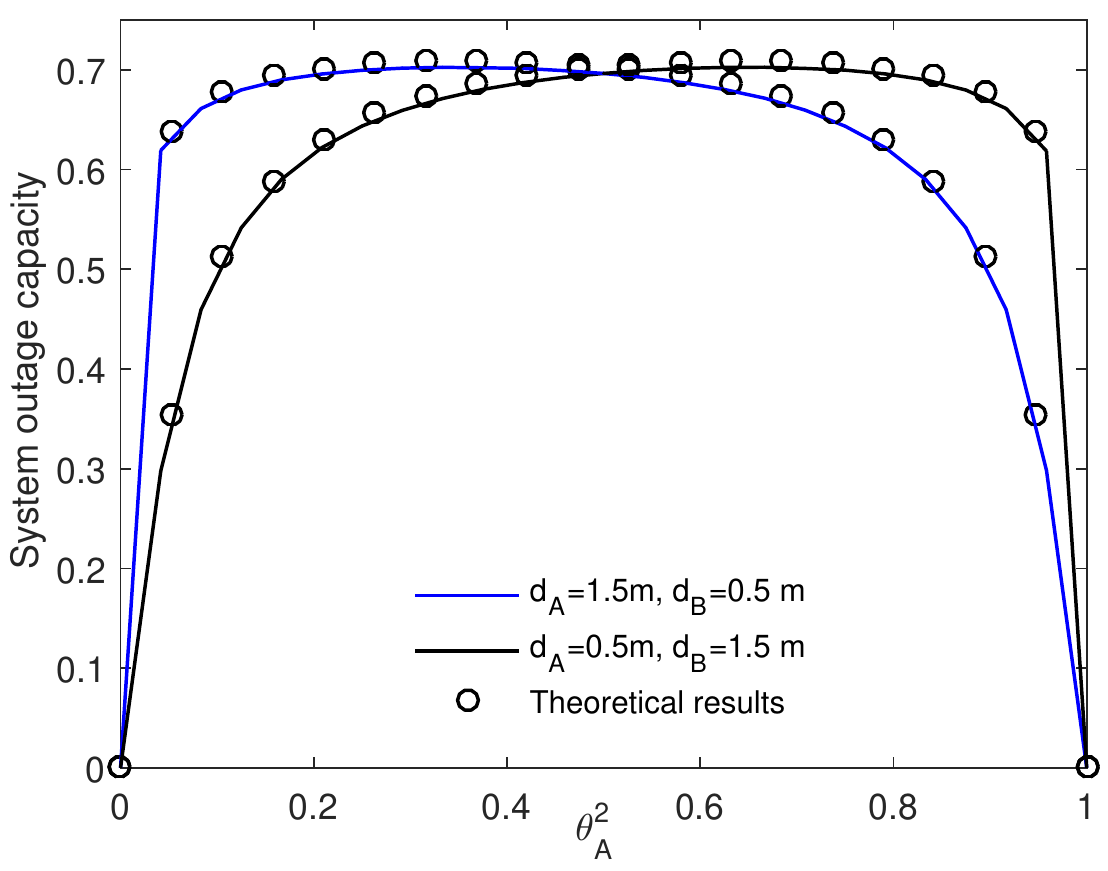}
\centering \caption{\color{black}{System outage capacity versus $\theta_{\rm{A}}^2$.}}
\end{minipage}
\end{figure*}
\subsection{Diversity Analysis}
The diversity gain is written as
\begin{align}
d =  - \mathop {\lim }\limits_{{\rho _0} \to \infty } \frac{{\log \left( {1 - P_{11}^s - P_{12}^s - P_{13}^s - P_{14}^s} \right)}}{{\log \left( {{\rho _0}} \right)}}.
\end{align}
It is not hard to verify that $\mathop {\lim }\limits_{{\rho _0} \to \infty } P_{11}^s = \mathop {\lim }\limits_{{\rho _0} \to \infty } P_{12}^s = \mathop {\lim }\limits_{{\rho _0} \to \infty } P_{14}^s = 0$ and  $\mathop {\lim }\limits_{{\rho _0} \to \infty } P_{13}^s = 1$. Thus, the diversity gain can be rewritten as
\begin{align}
d =  - \mathop {\lim }\limits_{{\rho _0} \to \infty } \frac{{\log \left( {1 - P_{13}^s} \right)}}{{\log \left( {{\rho _0}} \right)}}.
\end{align}
Now let us  compare  $\Omega _i$ and $\Phi _i$. When the transmit SNR approaches infinite, we have
\begin{align}
\mathop {\lim }\limits_{{\rho _0} \to \infty } \frac{{{\Omega _i}}}{{{\Phi _i}}} = \infty.
\end{align}
Combining (26) and the fact that $\Omega _i$ and $\Phi _i$ decrease with $\rho_0$, we have
\begin{align}
\mathop {\lim }\limits_{{\rho _0} \to \infty } \max \left( {{\Omega _i},{\Phi _i}} \right) = \mathop {\lim }\limits_{{\rho _0} \to \infty } {\Phi _i}.
\end{align}
Thus, the diversity gain can be calculated as
\begin{align}\notag
d &=  - \mathop {\lim }\limits_{{\rho _0} \to \infty } \frac{{\log \left( {1 - \exp \left( { - \frac{1}{{{\rho _0}}}} \right)} \right)}}{{\log \left( {{\rho _0}} \right)}}\\
&\mathop {{\rm{  }} = }\limits^{x = \frac{1}{{{\rho _0}}}} \mathop {\lim }\limits_{x \to 0} \frac{x}{{1 - \exp \left( { - x} \right)}} = 1.
\end{align}
\section{Simulations}

%\begin{figure}[htbp]
%  \centering
%    \includegraphics[width=0.3\textwidth]{fig3.pdf}
%  \caption{Outage capacity versus PS ratio.}
%\end{figure}
Here,  simulation results are provided to verify the derived expressions of \eqref{y5} and \eqref{y6}.
We adopt the same system parameter
settings as in \cite{2017CL} for  fair comparison between \eqref{y5} and the existing T2T outage capacity  \cite{2017CL}. %: $d_{\rm{A}}=d_{\rm{B}}=1 \rm{m}$, $\alpha=2.7$, $\mu _{\rm{A}}=\mu _{\rm{B}}=1$, $\beta=\frac{1}{3}$, $P_0=1.5$ J/s, $\sigma^2=0.01 \rm{W}$, $T=1$s, $U=3$ bit/s/Hz, and $\eta=1$. The unit of outage capacity is bit/Hz.
 {\color{black}Fig. 4 plots the relative approximate error versus $N$, as well as the outage capacity as a function of the  PS ratio by assuming that $\lambda_{\rm{A}}=\lambda_{\rm{B}}$ as in \cite{2017CL}.  For the upper part of Fig. 4, we assume that $\lambda_{\rm{A}}=\lambda_{\rm{B}}=0.2$. {\color{black}It can be seen  that the relative approximation error \cite{6828809} decreases with $N$, and that  our approximated results are sufficient to provide level of accuracy within very few items (e.g. $N\ge5$).}  For the lower part of Fig. 4, the upper and lower bounds of the T2T outage probability from \cite{2017CL} are also provided. The theoretical  T2T outage capacity and   system outage capacity with $N=5$ are computed based on \eqref{y5} and \eqref{y6}, respectively.}
{\color{black}It can be observed that the derived theoretical results match perfectly with the simulation results, which validates the correctness of the theoretical results. In particular, the derived theoretical results are accurate enough with $N=5$ no matter what the PS ratio is, which verifies our approximation again. Besides, the derived T2T outage capacity is more accurate than that in \cite{2017CL}, which is one of the contributions of this paper.} Another observation is that the achievable capacity increases first and then decreases with PS ratio, i.e., there only exists an optimal pair of symmetric PS ratio to maximize the achievable capacity.

 Fig. 5 plots the optimal system outage capacity and PS ratio as a function of the distance of $\rm{A}-\rm{R}$ link $d_{\rm{A}}$ with two PS schemes, respectively. The two PS schemes for comparison are the symmetric PS scheme and asymmetric PS scheme where both $\lambda_{\rm{A}}$ and $\lambda_{\rm{B}}$ are adjustable and the existing symmetric PS scheme with $\lambda_{\rm{A}}=\lambda_{\rm{B}}$ as in \cite{2017CL}. The  optimal  system outage capacity and PS ratios for two PS schemes are obtained by maximizing the derived system outage capacity expression, as in (23). It is difficult to derive closed-form solutions for the problem (23) by using conventional convex optimization methods. But we can use genetic algorithm-Based Algorithm  to solve (23)  (Please see Section 6 of \cite{jiang2017outage} as an example). For  convenience, we adopt the two-dimensional search algorithm to obtain the optimal PS ratios in Fig. 5.
It is assumed that  the relay is located on the straight line between  the both terminal
nodes, i.e.,  $d_{\rm{A}}+d_{\rm{B}}=2$. Note that here $d_{\rm{A}}\neq d_{\rm{B}}$ results in the unequal  statistic channel gains between the relay and the terminal
nodes.
One can see that the asymmetric PS scheme  provides a higher capacity compared with the symmetric PS scheme when $d_{\rm{A}}\neq d_{\rm{B}}$, while achieving the same capacity at $d_{\rm{A}}=d_{\rm{B}}$.  The reasons are as follows. The asymmetric PS scheme provides more flexibility than the symmetric PS scheme and the optimal PS ratios are determined by the statistic channel gains. If statistic channel gains are unequal, the asymmetric PS scheme makes two different optimal PS ratios and achieves a higher capacity; otherwise,
the optimal PS ratios of the asymmetric PS scheme satisfy $\lambda_{\rm{A}}=\lambda_{\rm{B}}$ and the PS ratio is equal to that of the symmetric PS scheme. One also can see that, in the asymmetric PS scheme, the relay  harvests more (or less) energy with a larger (or smaller) PS ratio from the terminal node with a better  (or worse) statistic channel to the relay, which achieves a trade-off between EH and information processing for the two terminal nodes.

Fig. 6 describes the relations of the optimal system outage capacity and  PS ratios against the  $\eta$, respectively. Here we set $d_{\rm{A}}$ and $d_{\rm{B}}$ as 0.8 and 1.2, respectively.
It can been seen that, with the increasing of  $\eta$, the optimal system outage capacity increases, while the optimal PS ratios decrease.  The reasons are as follows. Based on \eqref{B1} and \eqref{y6}, it is not hard to find that the maximum system outage capacity is to maximize {\small{$\min \left\{ {\mathop  \cup \limits_{i = {\rm{A}},{\rm{B}}} {\gamma _{i \to {\rm{R}}}} \ge {\gamma _{{\rm{th}}}},{\gamma _{{\rm{R}} \to i}} \ge {\gamma _{{\rm{th}}}}} \right\}$}}.  If $\eta$ increases,  ${\gamma _{{\rm{R}} \to \rm{A}}}$ and ${\gamma _{{\rm{R}} \to \rm{B}}}$ will increase. In order to maximize the system outage capacity, ${\gamma _{{\rm{A}} \to \rm{R}}}$ and ${\gamma _{{\rm{B}} \to \rm{R}}}$ should be improved, which can be realized by decreasing the optimal PS ratios. It can also be seen that the asymmetric PS scheme outperforms the symmetric one especially for a higher $\eta$.

In Fig. 7, we  plot  system capacity  versus  static power allocation ratio  $\theta_{\rm{A}}^2$ with different relay locations for $\lambda_A=\lambda_B=0.5$. {\color{black}It can be observed that the theoretical results match simulation results well, demonstrating the accuracy of our derived expression}. Besides, there exists an optimal $\theta_{\rm{A}}^2$ and the optimal $\theta_{\rm{A}}^2$ is  highly dependent on   the relay location. To maximize the   system capacity,  the relay should assign less power to $x_{\rm{B}}$  when the relay is close  to  terminal $\rm{A}$; otherwise, the relay should allocate more power to $x_{\rm{A}}$.  Finally, Fig. 8 verifies the diversity analysis.

\begin{figure}
  \centering
  \includegraphics[width=0.35\textwidth]{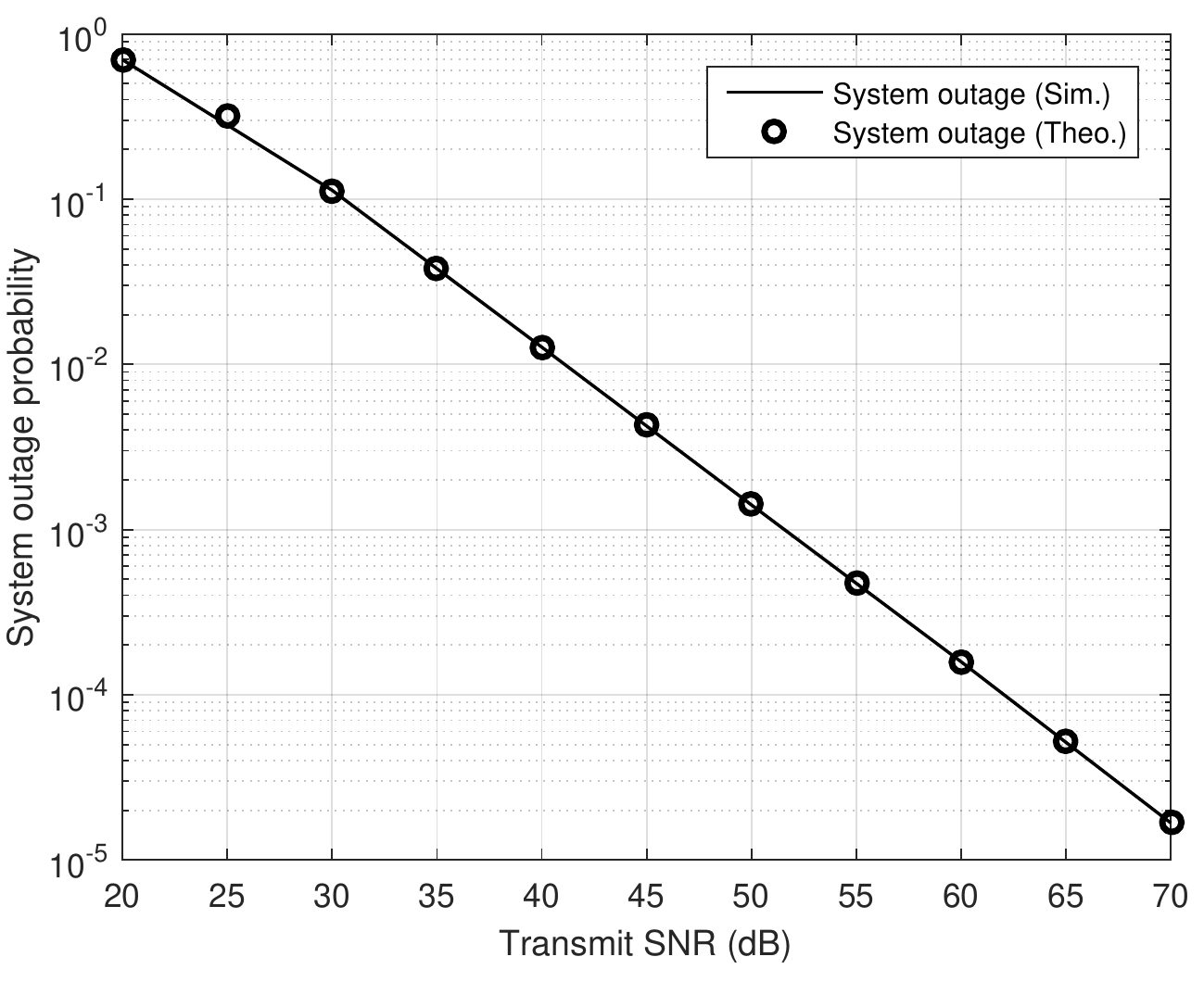}\\
  \caption{System outage probability against transmit SNR.}
\end{figure}
\section{Conclusions}
In this paper, we have studied the outage performance of  three-step DF TWRNs with PS SWIPT.   The expressions for T2T  and system outage probabilities (and capacities) have been derived and  verified by computer simulations. Some insights regarding the asymmetric PS scheme have been obtained. {\color{black}First,
the optimal PS ratios are sensitive to the relay location and the energy conversion efficiency. In particular, the optimal $\lambda_i$ decreases  with the increase of the statistic channel gain of $i$$-$$\rm{R}$ link. With the increase of the energy conversion efficiency, the optimal ratios, $\lambda_{\rm{A}}$ and $\lambda_{\rm{B}}$, show a downward trend. Second, the  asymmetric PS scheme outperforms the symmetric one for the unequal statistic channel gains between the relay node and the terminal nodes. Third, the selection of power allocation ratio is highly related with the relay location.}

{\color{black}This work can be extended by relaxing the assumption of perfect SIC, making our considered network close to practical scenario.}

%\vspace{-2em}
%\begin{figure}
%  \centering
%  \includegraphics[width=0.3\textwidth]{fig5.pdf}\\
%  \caption{Optimal $\tau_{\rm{out}}^{\rm{S}}$ and PS ratio  versus $\eta$.}
%\end{figure}
\ifCLASSOPTIONcaptionsoff
  \newpage
\fi
\bibliographystyle{IEEEtran}
\bibliography{refa}

\end{document}